\documentclass[sigconf]{acmart}

\newcommand*\cpp{C\kern-0.2ex\raisebox{0.4ex}{\scalebox{0.8}{+\kern-0.4ex+}}}
\usepackage{siunitx} 
\usepackage{cleveref}
\usepackage{enumitem}
\usepackage{framed}
\usepackage{listings}
\usepackage{multirow}
\usepackage{subcaption}

\usepackage[svgnames]{xcolor}

\newcommand{\enquote}[1]{``#1''}
\Crefname{lstlisting}{Listing}{Listings}

\AtBeginDocument{%
  }

\copyrightyear{2026}
\acmYear{2026}
\setcopyright{cc}
\setcctype{by}
\acmConference[ICSE-SEIP '26]{2026 IEEE/ACM 48th International Conference on Software Engineering}{April 12--18, 2026}{Rio de Janeiro, Brazil}
\acmBooktitle{2026 IEEE/ACM 48th International Conference on Software Engineering (ICSE-SEIP '26), April 12--18, 2026, Rio de Janeiro, Brazil}
\acmDOI{10.1145/3786583.3786919}
\acmISBN{979-8-4007-2426-8/2026/04}

\begin{document}



\title{On the Flakiness of LLM-Generated Tests for Industrial and Open-Source Database Management Systems}

\author{Alexander Berndt}
\email{alexander.berndt@uni-heidelberg.de}
\orcid{0009-0009-5248-6405}
\affiliation{%
  \institution{Heidelberg University}
  \country{Germany}
}
\additionalaffiliation{SAP}

\author{Thomas Bach}
\email{thomas.bach03@sap.com}
\orcid{0000-0002-9993-2814}
\affiliation{%
  \institution{SAP}
  \country{Germany}
}

\author{Rainer Gemulla}
\email{rgemulla@uni-mannheim.de}
\orcid{0000-0003-2762-0050}
\affiliation{%
  \institution{University of Mannheim}
  \country{Germany}
}

\author{Marcus Kessel}
\email{marcus.kessel@uni-mannheim.de}
\orcid{0000-0003-3088-2166}
\affiliation{%
  \institution{University of Mannheim}
  \country{Germany}
}

\author{Sebastian Baltes}
\email{sebastian.baltes@uni-heidelberg.de}
\orcid{0000-0002-2442-7522}
\affiliation{%
  \institution{Heidelberg University}
  \country{Germany}
}

\begin{abstract}
Flaky tests are a common problem in software testing.
They produce inconsistent results when executed multiple times on the same code, invalidating the assumption that a test failure indicates a software defect.
Recent work on LLM-based test generation has identified flakiness as a potential problem with generated tests.
However, its prevalence and underlying causes are unclear.
We examined the flakiness of LLM-generated tests in the context of four relational database management systems: SAP HANA, DuckDB, MySQL, and SQLite.
We amplified test suites with two LLMs, GPT-4o and Mistral-Large-Instruct-2407, to assess the flakiness of the generated test cases.
Our results suggest that generated tests have a slightly higher proportion of flaky tests compared to existing tests.
Based on a manual inspection, we found that the most common root cause of flakiness was the reliance of a test on a certain order that is not guaranteed (\enquote{unordered collection}), which was present in 72 of 115 flaky tests (63\%).
Furthermore, both LLMs transferred the flakiness from the existing tests to the newly generated tests via the provided prompt context.
Our experiments suggest that flakiness transfer is more prevalent in closed-source systems such as SAP HANA than in open-source systems.
Our study informs developers on what types of flakiness to expect from LLM-generated tests.
It also highlights the importance of providing LLMs with tailored context when employing LLMs for test generation.
\end{abstract}



\maketitle

\section{Introduction}
Large language models (LLMs) have emerged as a promising tool to automate test generation~\cite{llm-testing-survey}.
Although traditional test generation approaches, such as search-based techniques, have been shown to produce high-coverage test suites~\cite{pacheco2007feedback,harman2009theoretical}, the resulting test code suffers from poor readability and does not integrate well with existing tests~\cite{aster,sbst-combine-generation,sbst-combine2-generation}.
In contrast, LLMs have the ability to generate test code that appears natural to humans~\cite{sbst-combine-generation, pan2024assessing, hindle2016natural}.
Hence, various LLM-based test generation approaches have been evaluated~\cite{llm-harman-survey,llm-testing-survey}, leading to industrial adoption of such approaches~\cite{foster2025mutation,meta-generation}.

To evaluate the resulting tests, existing studies on LLM-based test generation typically measure the effectiveness of LLM-generated tests based on metrics such as code coverage and the mutation score, that is, quantifiable signals regarding the quality of the system under test~\cite{llm-testing-survey,mutation-generation}.
However, to our knowledge, there exists no dedicated research that focuses on the \emph{flakiness} of LLM-generated tests. 
Test flakiness is a major problem in industrial software testing, which has been studied by large software companies such as Google, Meta, Microsoft, and SAP~\cite{harman2018start,hoang2024presubmit,lam2020lifecycle,berndt2024taming}. 
Flaky tests produce inconsistent results when executed multiple times under the same conditions~\cite{luo2014empirical}. Such flaky behavior is undesirable as it reduces the efficiency and effectiveness of testing, thus leading to increased testing costs~\cite{memon2017taming,harman2018start,berndt2024test}.

Flaky tests invalidate the assumption that a test failure indicates the presence of a defect in the system under test.
Thus, developers may lose trust in the test suite and waste time debugging spurious failures~\cite{eck2019understanding}.
In continuous integration (CI) pipelines, flaky tests impede the automated assessment of code changes.
For example, Google reported that the CI pipeline in their testing infrastructure fails in 3.5\% of runs due to flaky failures~\cite{hoang2024presubmit}. 

Previous research on LLM-based test generation indicates that LLMs might produce flaky tests~\cite{pizzorno2024coverup,meta-generation}.
However, the prevalence of flaky tests and the root causes of flakiness remain unclear.
As a practical solution, previous work suggested repeatedly executing LLM-generated tests to filter flaky tests before handing them to developers for review~\cite{alshahwan2023software}.
Although this strategy can be helpful to filter out flaky tests, we consider it important to obtain additional information and a more holistic view of how and why LLM-generated tests are flaky~\cite{berndt2024taming}.
Gaining insight into the flaky tests that LLMs produce can help practitioners set realistic expectations about the types of flakiness they may encounter when reviewing and using LLM-generated test code~\cite{eck2019understanding,gruber2024automatic}.
Furthermore, comparing the prevalence of flakiness in LLM-generated tests with existing tests can help project the long-term flakiness rate of a test suite when LLMs are increasingly used to automatically generate tests.

In this work, we have evaluated the flakiness of LLM-generated tests for relational database management systems.
Previous research in the context of the large industrial database system SAP HANA has identified test flakiness as a major issue for automated testing~\cite{berndt2024taming,berndt2023vocabulary,berndt2024test,an2024just}.
Motivated by this previous work, we focused on test generation for SAP HANA.
In addition, we added three popular open-source databases to foster reproducibility and enable comparisons between open-source and closed-source software~\cite{baltes2025guidelinesempiricalstudiessoftware}. We consider this comparison valuable, as the source code of SAP HANA was not included in the training data, which might impact the results of our experiments.

We adapted Alshawan et al.'s approach for LLM-based test generation, which was previously applied in an industrial study at Meta~\cite{meta-generation}, to amplify tests in the four target systems and generate new test cases.
Our study focused on two types of tests: i) native unit tests written in \cpp~, and ii) SQL system tests.
We generated new tests for more than \num{5000} test files using two LLMs, GPT-4o and Mistral-Large-Instruct-2407~\cite{mistral-large,hurst2024gpt}.
In accordance with previous empirical studies on flakiness~\cite{parry2025systemic,berndt2024taming,gruber2024automatic,parry2023empirically}, we compiled and repeatedly executed the tests to obtain information on their flakiness.
We compared the prevalence and root causes of the resulting flaky tests based on a manual inspection of more than 100 samples.

Our results suggested that LLMs struggle to produce compilable \cpp~code, especially in the closed-source context of SAP HANA. We attribute this finding to the model's limited understanding of the provided code bases. In LLM-generated tests, the prevalence of flaky tests was often higher than in existing tests. 
Based on our manual inspection, we identified the reliance of tests on ordered test results from a collection that does not guarantee an order (\enquote{unordered collection}) as the most common root cause of flakiness in LLM-generated tests for database systems.

To test the robustness of LLMs against flakiness transfer from existing tests, we injected a flaky assertion into the original test code before providing it as input for generating additional tests.
In the generated tests, we found that both evaluated LLMs are susceptible to transferring the injected flakiness to generated tests.
Although LLMs transferred the injected flaky assertion to two-thirds of the newly generated SAP HANA tests, only half of the generated MySQL tests were affected.
Therefore, we conjecture that LLMs rely more heavily on the input context in a closed-source project, for which the source code was not present in the training data. Based on a discussion of these results with practitioners, we conclude that developers should prioritize fixing flaky tests before applying LLM-based generation approaches.

In summary, our study provides the following contributions.
We report (1) insights into the prevalence and root causes of flakiness in LLM-generated tests in the context of database management systems, 
(2) a comparison of the resulting flakiness between closed- and open-source database management systems and between two LLMs, and
(3) a benchmark on the susceptibility of LLM to transfer flakiness from the given context in a prompt to the generated tests.

We provide our code and datasets for the three open-source projects in a replication package~\cite{replication-package}.


\section{Background}
\label{sec:background}

In this section, we provide the flakiness definition that we adopted and flakiness categories motivated by prior work.

\subsection{Flakiness Definition}
Flaky tests produce inconsistent results, i.e., yield at least one \emph{positive} and one \emph{negative} result, when executed multiple times under the same conditions on the same state of the source code~\cite{gruber2024automatic}.
More formally, let $R_t = [r_1, ..., r_n]$ be the sequence of observed test executions for a test $t$ on the same state of the source code, where $r_i$ denotes the result of the $i$-th execution. Furthermore, let $r_i = 0$ for negative results and $r_i = 1$ for positive results. In accordance with previous work on flakiness, we considered a test $t$ flaky iff $0 < \sum_i^n r_i < n$, i.e., if the test did show both positive and negative results~\cite{berndt2024taming}.
We viewed \emph{passed} as the only positive result, and considered \emph{failed}, \emph{crashed}, or \emph{timed out} negative results.
We ignored \emph{skipped} executions. Tests were mainly skipped due to mismatched environmental preconditions, such as a specific operating system.

\subsection{Types of Flakiness}
Flaky tests can arise due to a variety of reasons. We focused on flakiness from the intra-flakiness family as introduced by \citet{parry2021survey}. We added \emph{uninitialized variable} and \emph{non-idempotent-outcome} (NIO) Flakiness as we considered them relevant for our experiments. This results in the following set of flakiness categories. 

\noindent\textbf{Unordered Collection.} Incorrect assumptions on a specific order in an inherently unordered data structure, such as a hash table~\cite{parry2021survey}. 

\noindent\textbf{NIO.} Side effects of a test execution persist in the execution environment, contaminating subsequent test executions~\cite{wei2022preempting}. 

\noindent\textbf{I/O.} Improper handling of file system operations, for example, when tests fail due to the disk running out of free space~\cite{parry2021survey}. 

\noindent\textbf{Uninitialized Variable.} Use of variables that have not been properly initialized~\cite{gruber2022survey,google-blog-flaky}. 

\noindent\textbf{Time.} Precision issues caused by improper use of system time~\cite{parry2021survey}. 

\noindent\textbf{Concurrency.} Unsafe handling of threads, leading to problems such as race conditions~\cite{parry2021survey}. 

\noindent\textbf{Randomness.} Use of random data generators without proper seeding, leading to inconsistent results~\cite{parry2021survey}. 

\noindent\textbf{Floating Point.} Inaccuracies in floating point operations, such as numerical overflows or underflows~\cite{parry2021survey}. 

\noindent\textbf{Too Restrictive Range.} Too narrow assertions, only allowing a subset of the valid outputs of the system under test~\cite{parry2021survey}.  

\noindent\textbf{Test Case Timeout.} Upper limit on execution time too low, causing intermittent test failures due to execution time variance~\cite{parry2021survey}.

\subsection{Flakiness of LLM-Generated Tests} 
In the following, we provide an overview of related work on LLM-based test generation that mentions flakiness as a potential problem. 

\citet{meta-generation} filtered flaky tests by repeatedly executing LLM-generated tests five times before presenting them to developers for review. 
In their experiments, 76\% of the Kotlin tests that built correctly passed reliably across five executions. However, this proportion also included tests that failed in all five executions, and the exact proportion of flaky tests remains unclear. 

\citet{pizzorno2024coverup} pointed out that LLM-generated tests can suffer from flakiness if the LLM is not provided with sufficient context. They report an example in which GPT-4o generated a flaky test that randomly assigns a value to the property of an object. Since the list of valid values of this property was unknown to the model, this random operation led to intermittent failures when the property was assigned an invalid value. However, since the main focus of their study was LLM-based test generation rather than flaky tests, they provided no information on how often such cases occurred or any other information on test flakiness.

In their framework for benchmarking LLMs for test generation, \citet{azanza2025tracking} introduced \emph{test isolation} as an important quality criterion of the resulting tests. Thereby, they aimed to prevent tests that are flaky due to a high reliance on external dependencies or due to inter-dependencies with other tests. Expert judges in the longitudinal study found that most LLMs are proficient in generating isolated tests, with GPT-4 being the only exception.

\section{Database Management Systems}
\label{sec:subjects}
The goal of our study was to gain insights into the flakiness of LLM-generated tests in the context of database management systems (DBMS). The observed types of flakiness may vary for other software projects~\cite{gruber2022survey}. We focused our study on two types of tests commonly encountered in database systems: native \cpp~tests~and SQL tests, which execute SQL statements against a running database~\cite{bach2022testing}. We chose SAP HANA and MySQL for our evaluation of native unit tests, as they both share GoogleTest as one of the adopted test frameworks. 
We further examined DuckDB's and SQLite's SQL tests, as they represent lightweight alternatives offering short build and test times. \Cref{tab:testcases} provides a summary of the considered DBMS and included test cases.

\subsection{SAP HANA}
SAP HANA is an in-memory DBMS that was introduced by SAP in 2010~\cite{bach2022testing}. 
Typically, SAP software is used to orchestrate business operations, often using SAP HANA to store and retrieve business-critical data~\cite{jambigi2022automatic}. 
Thus, failures of SAP HANA can have a negative impact on customers' business operations. 
This, in turn, justifies high investments in an extensive and continuous testing process.

SAP HANA's source code consists of millions of source lines of code (SLOC), mostly written in \cpp~\cite{jambigi2022automatic, berndt2024taming}. 
The regression test suite of SAP HANA mainly includes two types of tests: 1) native \cpp~unit tests that verify a certain functionality in isolation and 2) system tests that execute SQL statements on a running database~\cite{bach2022testing}.

Due to the large amount of resources required for testing, SAP utilizes a multistage approach to test SAP HANA~\cite{berndt2024taming}. The different testing stages vary in resource demands and the frequency with which the respective tests are executed. These stages range from the local testing of developers in the initial stage to extended manual and automated testing for final release qualification. To cope with the resource demands of the tests executed in the later testing stages, SAP has set up a central testing infrastructure where tests are executed in parallel on over \num{1000} servers. However, although this central testing infrastructure offers vast computational resources for test execution, previous work has pointed out that such complex testing environments may also cause flakiness, e.g., through hardware failures~\cite{berndt2024taming,pfs-facebook,silva2024effects}. 
In this work, we evaluated the flakiness of DBMS tests, which, according to our definition in \Cref{sec:background}, requires repeated test executions in a controlled environment~\cite{gruber2024automatic}. To minimize the flakiness caused by environmental factors during repeated test executions~\cite{silva2024effects}, instead of relying on the central testing infrastructure, we conducted the test executions on a dedicated server. As executing tests from the later testing stages of SAP HANA would lead to long execution times, we focused on tests that are used for local developer testing, i.e., native \cpp~tests, written with the GoogleTest framework~\cite{googletest-docs}. 

\subsection{MySQL Server}
MySQL Server is an open-source DBMS initially released in 1995~\cite{mysql-github}. 
Today, Oracle offers MySQL Server in two versions: the open-source MySQL Community Server and the proprietary MySQL Enterprise Server.
For this work, we examine the open-source MySQL Community Server (in the following referred to as MySQL).

The source code in MySQL's GitHub repository~\cite{mysql-github} is mostly written in C/\cpp~and consists of approximately 7 million SLOC, measured using CLOC~\cite{cloc}.
Similar to SAP HANA, the MySQL test suite mainly comprises two types of tests: SQL tests written in a dedicated test language and unit tests written in \cpp. Also similar to SAP HANA, MySQL has a set of unit tests that are written using the GoogleTest framework. To enable a comparison between the results for SAP HANA and MySQL, we examined these GoogleTest unit tests in our study. 

\subsection{SQLite}
SQLite is a lightweight open-source DBMS mostly written in C~\cite{sqlite-repo}.
According to its documentation, SQLite is the most widely used database engine in the world~\cite{sqlite-deployed}.
The main use case of SQLite is to provide a local data storage for applications that is easy to use and requires no complex setup or administration. 

Compared to SAP HANA and MySQL, the SQLite repository contains fewer lines of code, comprising approximately \num{400000} SLOC. 
For testing the SQLite core library, there exist four different test harnesses: TH3, TCL, SQL Logic Test, and dbsqlfuzz~\cite{sqlite-deployed}.
TH3 and dbsqlfuzz are proprietary harnesses and are not publicly available. 

Therefore, in our study, we focused on SQLite's Tool Command Language (TCL) test harness, which is publicly available and serves as the main test suite for testing SQLite during development.  
The TCL test harness provides a TCL wrapper around SQL tests that are executed against a running database instance, similar to SAP HANA's system tests.
Thus, our goal was to gain insights into the flakiness of LLM-generated tests that contain SQL statements.
 
\subsection{DuckDB}

DuckDB is a lightweight analytical RDBMS developed in \cpp~\cite{duckdb-docs}.
Since its release in 2018, DuckDB has become an emerging DBMS that is continually refined~\cite{muhleisen2025runtime,kuiper2024robust,kuiper2023these}.
With its vectorized query execution engine, DuckDB is specifically optimized for online analytical process (OLAP) workloads.

DuckDB's source code is mostly written in \cpp, comprising almost 2 million SLOC. 
The DuckDB test suite consists of native \cpp~tests written with the Catch2 unit testing framework and SQL tests written with sqllogictest~\cite{duckdb-docs}. 
While native \cpp~tests are specifically targeting the \cpp~API of DuckDB, DuckDB's developers aim to test most functionality with the help of SQL tests.
For our study, we focused on DuckDB's SQL tests to gain insights into the flakiness of LLM-generated SQL tests.

\begin{table}[tb]
  \caption{The number of relevant test files and test cases in the original test suites of the study subjects. Note that (1) the number of relevant files and tests represents only a subset of SAP HANA's unit tests, which we selected with the assistance of practitioners, and (2) the numbers depend on the different ways in which the test frameworks count tests.}
  \centering
  \small
  \begin{tabular}{l l r r}
    \toprule
    Language & Project & \#Test Files & \#Tests \\
    \midrule
    \multirow{2}{*}{\cpp} 
    & MySQL   & \num{260} & \num{2127} \\
    & SAP HANA & \num{1525} & \num{19947} \\
    \midrule
    \multirow{2}{*}{SQL}
    & DuckDB  & \num{2748} & \num{236185} \\
    & SQLite  & \num{603} & \num{388282} \\
    \bottomrule
  \end{tabular}
  
  \label{tab:testcases}
\end{table}

\begin{figure*}
  \includegraphics[width=0.85\textwidth]{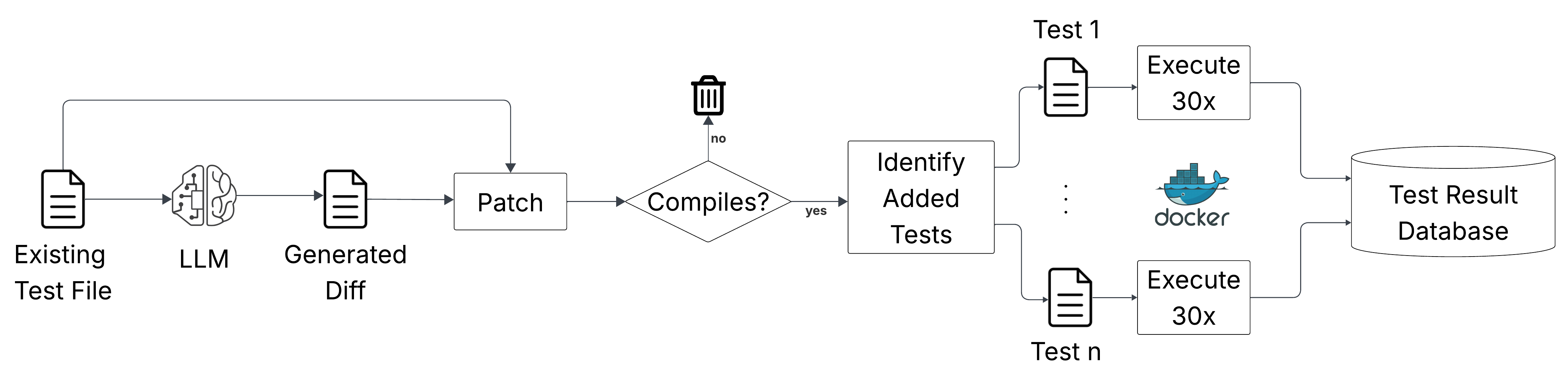}
  \caption{Study setup inspired by \citet{meta-generation}: We first prompt the LLM to generate new tests given an existing test file. Then, the generated code is added to the existing test file before compilation. After compilation, we identify the newly added, generated tests, execute them in separate Docker containers, and store the results in a database.}
  \Description{A diagram illustrating the approach described in \Cref{sec:methodology}.}
  \label{fig:approach}
\end{figure*}
\section{Methodology and Research Questions}
\label{sec:methodology}
In this section, we motivate our research questions and outline our methodology for addressing them.

\subsection{RQ1: Prevalence}
With the first research question, we wanted to gain insights into the prevalence of flakiness in LLM-generated tests.
\begin{quote}
    \textbf{RQ1}: \emph{What is the proportion of flaky tests in LLM-generated tests for test amplification?}
\end{quote}

\textbf{Motivation:} Automated test generation (ATG) is one of the most commonly targeted tasks by previous research on LLM-based software testing~\cite{llm-testing-survey}. Thus, many potential approaches to employ an LLM for test generation have been proposed~\cite{meta-generation,foster2025mutation,mutation-generation,dakhel2024effective,aster,yang2025kernelgpt,a3test,automation-generation,chatunitest,sbst-combine-generation}, most of them focusing on the generation of regression tests~\cite{harman2025harden}. The goal of this research question was to gain insights into the prevalence of flakiness in LLM-generated regression tests. With this, we aimed to project the long-term impact of an increasing use of LLM-based test generation on the flakiness of a test suite. Existing LLM-based ATG approaches typically utilize LLMs to generate the desired test code directly~\cite{a3test}, or in combination with traditional ATG tools such as Pynguin or Evosuite~\cite{sbst-combine-generation,fraser2014evosuite,lukasczyk2022pynguin,kitsios2025automated}. In this study, we chose to employ a standalone LLM approach, as the scalability of traditional ATG tools to large code bases is limited~\cite{huang2025benchmarking}.
Moreover, we did not find any readily available solution for our study subjects, that is large \cpp~projects.

\textbf{Approach:}
We adopted an approach of Alshawan et al.~\cite{meta-generation}, which was previously deployed at Meta with promising results. \Cref{fig:approach} illustrates our approach, which we chose for three main reasons: i) it was already tested in a large industrial software project with promising results, ii) it amplifies existing test suites and is therefore tailored for projects that already contain a large number of existing tests, and iii) it provides the existing test code as an input for an LLM, which we expected to increase the chance of obtaining compilable and executable output for closed-source projects with highly customized and rather complex test code such as SAP HANA. Furthermore, the chosen approach utilizes a relatively simple setup to generate flaky tests, which enabled its adoption in the context of SAP HANA. In the following, we detail our setup to generate tests for our flakiness evaluation.

\textbf{Prompt}. In accordance with \citet{meta-generation}, we instructed the model to \enquote{Write an extended version of the test file that includes additional tests to cover some extra corner cases.}. We provided the existing test file as input context, as previous work has shown that LLMs are capable of generating tests that amplify the current test suite given a minimal context~\cite{meta-generation}. We chose this rather simple setup, as the focus of our study was only to generate a sample of tests for our flakiness evaluation, assuming that flakiness is evenly distributed between tests generated by different approaches for LLM-based ATG. Finally, we added an output indicator to the prompt so that we could parse the generated code from the LLM's response and patch the existing test file. 

\textbf{Models.} We conducted our experiments using two LLMs, GPT-4o and Mistral-Large-Instruct-2407 (in the following: Mistral). 
Both models were provided by an internal API at SAP, which can be used with an SDK that supports LangChain integration~\cite{langchain}, allowing up to 100 requests per minute. 
We mainly chose GPT-4o because it was the best-performing model~\cite{abdullin2025testwars} available via the internal API, which allowed us to use internal data (i.e., the SAP HANA source code). We repeated our experiments with Mistral's large instruct model to foster the reproducibility of the results for the open-source projects in this study in accordance with existing guidelines on empirical studies using LLMs~\cite{wagner2025towards, baltes2025guidelinesempiricalstudiessoftware}. We chose Mistral as the open-source alternative as it was the only open-source model available via the internal API at the time of our study. We set the temperature to zero for both models, as this setting has been shown to produce tests with the highest compilation success rate (CSR) in a previous study~\cite{meta-generation}. Both models have a context window of 128k tokens. While GPT-4o's number of parameters is estimated at 200b~\cite{abacha2024medec}, Mistral-Large-Instruct-2407 has 123b parameters.

Based on the given prompt and models, we generated code for each of the files listed in \Cref{tab:testcases} using both LLMs. After parsing the newly generated test code from the LLMs' response, we extended the original test files with the new code. By integrating the generated code into existing test files, we were able to utilize the test frameworks of the respective projects as-is, which simplified the automated execution of the resulting test code. After patching the original test files, we compiled the resulting code to filter syntactically incorrect files. 

\textbf{Execution Environment.}
To ensure a clean environment for every test execution and avoid flakiness caused by environmental factors, we executed the tests in separate Docker containers, which is common practice in empirical studies on test flakiness~\cite{gruber2024automatic,silva2024effects,lam2019idflakies,parry2023empirically,wang2022ipflakies,zhang2021domain,rahman2024flakesync}. For this, we created a dedicated Docker image for each of the projects in our study. These Docker images contained a compiled version of the respective project, so after launching a container based on the image, we only had to patch the relevant test file before initiating the test execution. We report the commit hashes of the projects we used for this study in our replication package. In the following, we briefly describe the execution setup for the two types of tests in this study. 

\textbf{Executing \cpp~Tests.} For the \cpp~tests, we launched a Docker container for each \emph{test} before executing the tests one-by-one in isolation. In preliminary experiments, we found that \enquote{naively} executing generated tests for database systems at scale may lead to crashes of the execution server. For example, a generated test for SAP HANA's memory management tried to allocate a very large amount of memory, thus causing resource contention. We limited the impact of such cases by providing the Docker containers with a fixed amount of computational resources and timing out long-running test executions after five minutes. Note that this specific execution environment also sets limits on the types of flakiness that we could possibly observe. For example, by providing the test executions with a fixed amount of computational resources, we expected to reduce the impact of resource-affected flaky tests~\cite{silva2024effects}. More specifically, we expected mainly observe intra-test flakiness in our experiments, as our test isolation via Docker minimizes the probability of failures caused by inter-dependencies between tests~\cite{parry2021survey}. We made this decision to increase the control and focus of our study and to identify only the types of flakiness that were caused by the generated code itself.

\textbf{Executing SQL Tests.} In contrast to the \cpp~tests, we launched the Docker containers for the SQL tests for each \emph{test file}. 
This is due to the structure of the SQL tests. The SQL tests are order-dependent by design, i.e., the custom test frameworks of SQLite and DuckDB always execute the tests in a file in the same order, thus explicitly allowing developers to write order-dependent tests. \Cref{lst:sql-tests} illustrates an example from DuckDB. In this case, the test aims to test the \texttt{join} operator. For this, a table is first created and populated, before the \texttt{join} operator is applied. Although one might consider such tests inherently flaky, as executing them in a different order would cause test failures, we emphasize that the focus of our study was on practically relevant flakiness. Therefore, we executed the tests in an environment that resembles the testing environment in practice, where the execution order is predetermined.

\textbf{Existing Tests.}
To ensure that our execution environment meets the requirements for the given projects and to identify any potentially existing flaky tests, we first executed all relevant existing tests from the different projects $n=30$ times and stored the results from these test executions in a database for later analysis. We chose to perform 30 repeated executions, as our goal was to detect flaky tests that are practically relevant, i.e., tests that fail with a high probability.

\subsection{RQ2: Root Causes}
With the second research question, our aim was to find the most prominent root causes of test flakiness in LLM-generated tests.

\begin{quote}
    \textbf{RQ2}: \emph{What are the flakiness root causes of LLM-generated tests for test amplification?}
\end{quote}

\textbf{Motivation:} One might expect that the root causes for the flakiness of LLM-generated tests follow the distribution of existing tests, as LLMs essentially mimic human behavior they learned from their pre-training. However, as discussed in \Cref{sec:background}, previous research suggests that flakiness in LLM-generated tests can occur due to different root causes~\cite{pizzorno2024coverup,azanza2025tracking}. For example, LLMs may generate flaky tests due to a lack of sufficient context, which is available to developers but has not been externalized in the context of a prompt. Furthermore, previous work also pointed out that the root causes of flakiness vary between different types of software~\cite{gruber2022survey}. Therefore, we investigated the root causes of flaky tests generated by LLMs in the context of DBMS.

\textbf{Approach:} To gain insights into the root causes of flakiness in LLM-generated tests, we manually labeled the flaky tests identified based on our definition of flakiness in \Cref{sec:background}, identified by repeated test executions. 
We conducted the labeling using the test results of the repeated test executions, the resulting error messages in failing executions if reproducible, and the executed test code. 
In accordance with previous research on flakiness~\cite{gruber2024automatic}, we utilized the categories by \citet{parry2021survey} as described in \Cref{sec:background} for our labeling.
To enable a comparison between the root causes of flakiness for tests written by developers and those generated by the LLM, labeling was conducted for the flaky tests resulting from the generation experiments, as well as for existing flaky tests. 

\subsection{RQ3: Transfer}
Our third research question concerned the degree to which LLMs are susceptible to transferring flakiness from existing tests to the newly generated tests.
\begin{quote}
    \textbf{RQ3}: \emph{To what degree does an LLM transfer flakiness present in the provided in-context example tests to the newly generated tests?}
\end{quote}

\textbf{Motivation:} Previous work on LLM-based test generation found that when generating new tests, LLMs follow existing coding conventions and adhere to the structure of the code presented in the context of a prompt~\cite{foster2025mutation,meta-generation}.
\citet{harman2025harden} called this effect \enquote{Fashion Following} and argued that the resulting test code is thus more intuitive to human readers. 
This \enquote{Fashion Following} can be attributed to the ability of LLM to learn from examples in the given input context, that is, in-context learning (ICL)~\cite{brown2020language,wei2022emergent}. ICL has been shown to be an effective approach to improve the performance of an LLM on various tasks~\cite{brown2020language}. 
However, previous research has also shown that LLMs lack robustness against spurious correlations when learning from the provided context~\cite{tang-etal-2023-large,hermann2023foundations}. We hypothesized that, for test amplification, this may lead to the transfer of flakiness from the existing tests, which are shown to the LLM, to the newly generated test code. Therefore, we aimed to gain insights into the robustness of LLMs against such a flakiness transfer. 

\textbf{Approach:} We injected flakiness into existing tests and replicated our test generation experiment with the injected flaky tests. More specifically, we added an assertion to the \cpp~tests that fails with 50\% probability, using the following steps~\cite{cordy2022flakime}: 
(1) Parse the original code of the given tests from the \cpp~files.
(2) Add an assertion to the test, which may fail based on the outcome of a coin flip at the end of the test code.
(3) Replace the original test code with the injected code for every test in all relevant test files.
(4) Persist the injected files for later use.
We performed these steps before providing the test files to the LLM and evaluated whether the LLM \enquote{follows} the injected flakiness in its generated output.

\subsection{Threats to Validity}
\label{sec:threats}

\begin{figure*}
  \centering
  \begin{subfigure}{0.49\textwidth}
    \includegraphics[width=\textwidth]{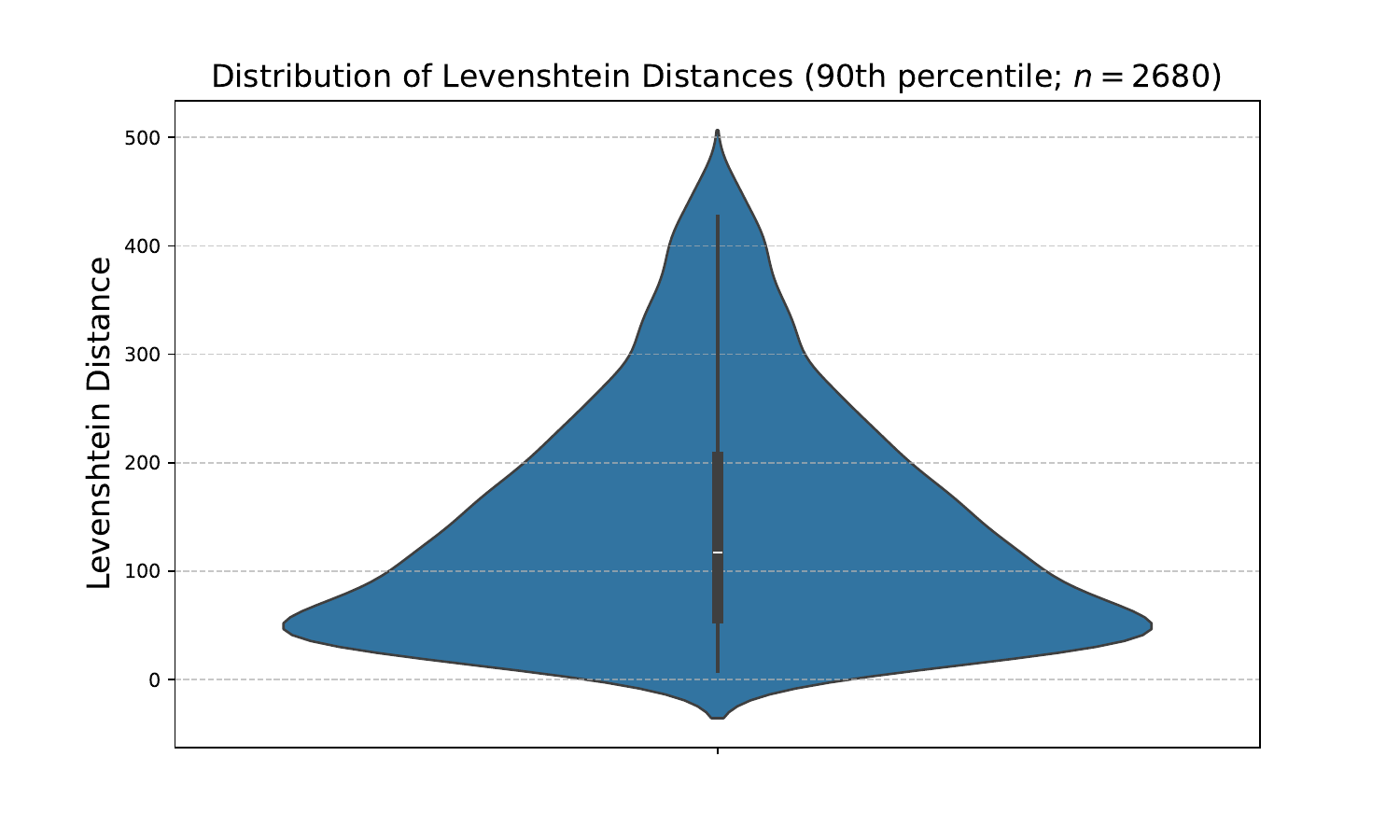}
    \caption{Tests generated by GPT-4o using existing tests in SAP HANA.}
  \end{subfigure}
  \begin{subfigure}{0.49\textwidth}
    \includegraphics[width=\textwidth]{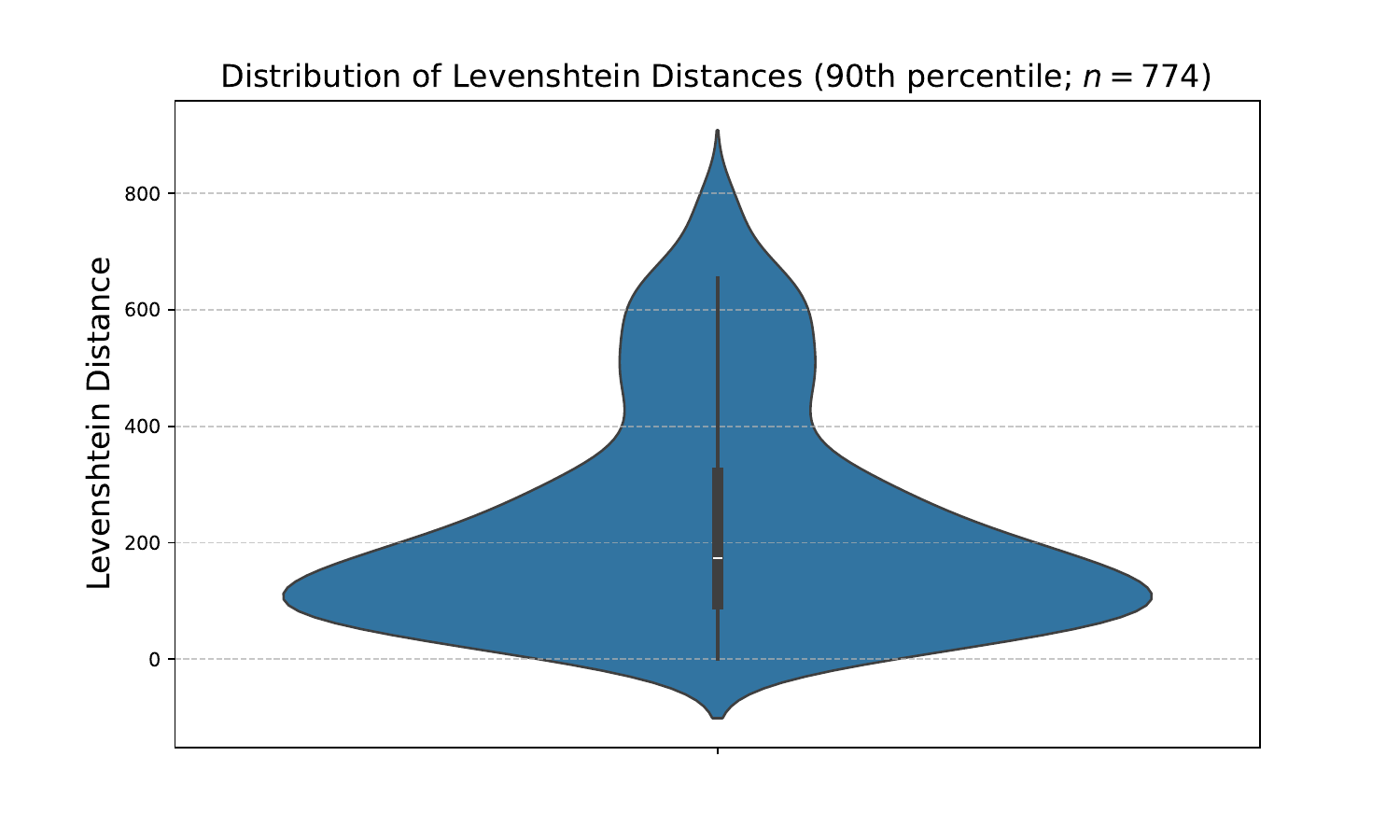}
    \caption{Tests generated by GPT-4o using existing tests in MySQL.}
  \end{subfigure}
  \caption{Violin plots displaying the syntactical difference between generated and existing tests.}
  \Description{Distributions of LEDs for syntactical comparisons.}
  \label{fig:clones}
\end{figure*}

\textbf{Internal Validity.}
In contrast to \citet{meta-generation}, we did not calculate the code coverage of the resulting tests. Instead, we kept all compiling tests for our flakiness evaluation. To mitigate the threat of analyzing mostly nonsensical tests, we conducted a qualitative analysis of the flaky tests during our manual inspection of flakiness categories. Furthermore, we analyzed the syntactic similarity of the generated tests with the existing tests to ensure that the LLM did not simply reproduce existing tests with new inputs at scale. We measured the syntactic similarity with the Levenshtein Edit Distance (LED), as it is a simple but effective measure to compare the syntactical similarity of generated test code~\cite{ouyang2025empirical}. To achieve this, we compared each existing test with each newly generated test in a file. For example, in a file with $n=3$ existing tests and $m=2$ LLM-generated tests, we compared both LLM-generated tests with the 3 existing tests, thus ending up with $n \times m = 6$ comparisons. We report the distribution of the closest distances between existing and LLM-generated tests per file in \Cref{fig:clones}.

Since LLMs are inherently non-deterministic, multiple repetitions of the same experiment may lead to different results. To mitigate the threat of the inherent non-determinism of LLM, we followed two approaches. First, we repeated relevant parts of our experiments to measure the degree to which our conclusions could be impacted by randomness. We performed these repetitions with GPT-4o, as it consistently yielded better results than Mistral. More specifically, we repeated the generation for files containing flaky tests in our initial experiments to see whether the flakiness persists, and repeated the injection experiments two additional times to ensure the reported proportion of tests containing the flaky assertion is valid. Furthermore, in all our experiments, we set the temperature to zero, as this setting has been shown to promise the most consistent results in a previous study~\cite{ouyang2025empirical}.

\textbf{Construct Validity.} In contrast to previous studies on flakiness, which have performed up to \num{10000} repeated executions of a test to gain insights into its flakiness~\cite{alshammari2021flakeflagger}, we executed the tests in this study only 30 times. Since some flaky tests have been shown to fail rarely~\cite{berndt2024taming}, we may therefore underestimate the prevalence of flakiness in our study. For example, assuming that test executions are independent and identically distributed, we expected to identify tests with a failure probability of 15\% with a probability of 99\%. With \num{10000} repeated executions, tests with a failure probability of 0.04\% could be identified with 99\% probability. However, using 30 repeated executions to identify flaky tests is similar to the reality of SAP's approach for detecting flaky tests, which is based on the trade-off between detecting all flaky tests and overestimating the flakiness due to environmental noise. 

\textbf{External Validity.} Our study setup is limited to a single LLM-based ATG approach in the idiosyncratic context of relational database systems. Therefore, the resulting flakiness distributions we observe might not generalize to other contexts, such as NoSQL database systems or different software domains. Furthermore, given the small number of flaky tests in our study, the statistical significance of our results is limited. We view the results of our study as a first step toward more detailed insights into the flakiness of LLM-generated tests. We motivate other researchers to experiment with different prompts, LLM-based ATG setups, and software projects.

\begin{lstlisting}[
  basicstyle=\ttfamily\scriptsize,
  numbers=left,
  numbersep=6pt,
  stepnumber=1,
  showstringspaces=false,
  breaklines=true,        
  breakatwhitespace=true, 
  postbreak=\mbox{\textcolor{red}{$\hookrightarrow$}\space}, 
  keywordstyle=\color{blue}\bfseries,
  commentstyle=\color{DarkGreen}\itshape,
  columns=fullflexible,
  keepspaces=true,
  frame=single,
  language=SQL,
  caption={Example structure of an SQL test for DuckDB.}, 
  captionpos=b, 
  label={lst:sql-tests},
  float
]
statement ok
CREATE TABLE test (a INTEGER, b INTEGER);
statement ok
INSERT INTO test VALUES (11, 1), (12, 2), (13, 3);
statement ok
CREATE TABLE test2 (b INTEGER, c INTEGER);
statement ok
INSERT INTO test2 VALUES (1, 10), (1, 20), (2, 30);
# simple cross product + join condition
query III
SELECT a, test.b, c FROM test, test2 WHERE test.b = test2.b ORDER BY c;
----
11	1	10
11	1	20
12	2	30
\end{lstlisting}

\section{Results}
\label{sec:generation-results}
In this section, we report the results of our experiments. 
We start by giving an overview of the execution results of existing tests.
Afterwards, we report the results of the LLM-generated tests.

\begin{table}
  \caption{An overview of the relevant existing tests in our study subjects.}
  \label{tab:original} 
  \centering
  \small
  \begin{tabular}{l r r r r}
    \toprule
     Metric & SAP HANA & MySQL & SQLite & DuckDB \\
    \midrule 
    \#Tests & \num{19947} & \num{2127} & \num{388282} & \num{236185} \\
    \#Flaky Tests & 13 (0.07\%) & 13 (0.61\%) & 0 & 0 \\
    \bottomrule
  \end{tabular}
\end{table}

\begin{table*}
  \caption{The results from the executions of our LLM-generated tests.  Note that SQLite and DuckDB tests are written in SQL and do not need to be compiled (compilation success rate marked with \enquote{---} in the table).}
  \label{tab:results}
\small
  \centering
    \begin{tabular}{l l r r r r r}
      \toprule
      & Project & \#Diffs & \#Compilation Successes & \#Added Tests & \#Passing Tests & \#Flaky Tests \\
      \midrule
      \multirow{4}{*}{GPT-4o} 
      & SAP HANA & \num{1525} & 707 (46\%) & \num{2975} & \num{1678} (56\%) & 5 (0.29\%) \\
      & MySQL & 260 & 173 (67\%) & 868 & 550 (63\%) & 1 (0.18\%) \\
      & SQLite & 598 & --- & \num{133592} & \num{126774} (95\%) & 4 (0.00\%) \\
      & DuckDB & \num{2688} & --- & \num{71031} & \num{57591} (81\%) & 42 (0.07\%) \\
      \midrule
      \multirow{4}{*}{Mistral} 
      & SAP HANA & \num{1525} & 616 (40\%) & \num{3257} & \num{1773} (54\%) & 6 (0.39\%) \\
      & MySQL & 241 & 138 (57\%) & 861 & 560 (65\%) & 4 (0.71\%) \\
      & SQLite & 592 & --- & \num{60123} & \num{52276} (87\%) & 6 (0.01\%) \\
      & DuckDB & \num{2676} & --- & \num{119666} & \num{100794} (84\%) & 47 (0.05\%) \\
      \bottomrule
    \end{tabular}
\end{table*}

\subsection{RQ1: Proportion of Flaky Tests}
In our first research question, our aim was to gain insights into the prevalence of flakiness in LLM-generated tests. 
In the following, we report the quantitative results of our test executions.

\textbf{Existing Tests.}
\Cref{tab:original} shows the results of 30 repeated executions of the existing tests.
We observed that both \cpp~projects show a small number of flaky tests. 
For MySQL, 13 of \num{2127} (0.6\%) passing tests show flaky behavior. 
This number is similar to the flakiness numbers reported in previous work~\cite{gruber2024automatic}.
Analyzing the 13 flaky tests, we observed that they originated from four different test files. They all suffered from NIO flakiness, i.e., the tests self-pollute the execution environment in their first execution.
\Cref{lst:nio-example} shows an example of an NIO-flaky MySQL test.
In the example in \Cref{lst:nio-example}, the test uses static variables that are not reinitialized in subsequent executions, thus causing the assertions in lines 18 and 19 to fail.

We observed no flaky tests in the existing tests of SQLite and DuckDB.
While all original tests for DuckDB yielded a passing result, ten tests of SQLite did not pass in our execution environment.
We found that one of these tests exceeded our five-minute timeout, one was skipped, and eight tests failed.
All failing tests resulted from a single file containing tests for SQLite's write-ahead logging (WAL) mode~\cite{sqlite-deployed}.
Each of these tests required ten threads simultaneously, which collides with our approach to parallelized test execution. We omitted this file in the following experiments.

\textbf{Generated \cpp~Tests.}
\Cref{tab:results} shows the execution results of the LLM-generated tests.
Using GPT-4o, we generated \num{1785} diffs resulting in \num{3843} added tests.
For Mistral, we obtained a similar number with \num{4012} added tests.
The compilation success rate (CSR) differed between MySQL and SAP HANA both for Mistral and GPT-4o.
While GPT-4o produced compilable code for two out of three files for MySQL,
less than half of the resulting files for SAP HANA were compilable. 
We consider this result intuitive, as SAP HANA's source code was not part of the training data 
and previous research suggests that missing background knowledge on the project context may cause compilation errors due to hallucination~\cite{zhang2025llm}.
Comparing Mistral to GPT-4o, we observed a slightly lower CSR for Mistral, with a drop by a factor of 0.87 for SAP HANA and 0.82 for MySQL. 
Similarly, comparing the pass rates for SAP HANA and MySQL, we observed that tests generated for MySQL yield a slightly higher pass rate (63\% vs. 56\%) for both LLMs.

We observed a relatively low number of flaky tests for both SAP HANA and MySQL.
GPT-4o generated a single flaky test for MySQL. For SAP HANA, we observed flaky behavior in five generated tests. 
With six flaky tests for SAP HANA and four for MySQL, Mistral generated a higher number of flaky tests, resulting in a slightly higher proportion of flaky tests in both projects. 

\textbf{Generated SQL Tests.}
We report the results of the SQL tests in \Cref{tab:results}.
Compared to the two \cpp~projects, we observed higher pass rates for SQL tests, ranging between 81\% and 95\%.
Note that there was a large difference in the number of generated tests for DuckDB and SQLite between models. We attribute this difference to the way we count SQL tests. A single SQL test in the code may be executed for different configurations of the database on which the test is executed. Since we counted each instance of such an execution as a distinct test, one additional generated test may increase the number we count by more than one. 
Although there were no existing flaky tests in the original test suites of both projects, both models generated flaky tests for SQLite and DuckDB. GPT-4o generated a relatively low number of flaky tests for SQLite (4), but a notably higher number of flaky tests for DuckDB (42). With six flaky tests for SQLite and 53 for DuckDB, Mistral generated more flaky tests than GPT-4o, similar to our results for the \cpp~tests.

\textbf{Syntactical Similarity.} \Cref{fig:clones} displays the LEDs between existing and LLM-generated tests. We observed an average LED of 299.14 between LLM-generated and existing tests for SAP HANA (min=8, $Q_1$=62, $Q_3$=250, max=\num{45926}). Furthermore, for SAP HANA, there was no LLM-generated test that was an exact copy of an existing test. 
For MySQL, the syntactic similarity between LLM-generated and the closest existing test per test file is higher, with an average of 405 (min=0, $Q_1$=101, $Q_3$=466, max=\num{10154}).
Overall, we conclude that the resulting LEDs suggest a sufficiently large syntactical dissimilarity between existing tests and tests generated by the LLMs.

\begin{framed}
\noindent\textbf{Result RQ1 (Proportion):} The proportion of flaky tests generated by the LLM varied between projects and models. Compared against existing tests, we observed a higher proportion in the generated \cpp~tests by GPT-4o for SAP HANA. For MySQL, the proportion of flaky tests was lower for the generated tests. For Mistral, the proportion of flaky \cpp~tests is slightly higher. Although we found no flaky SQL test written by a developer, we found 96 flaky tests in the tests generated by the LLMs in this study (43 for GPT-4o, 53 for Mistral).
\end{framed}

\begin{lstlisting}[
  basicstyle=\ttfamily\scriptsize,
  numbers=left,
  numbersep=6pt,
  stepnumber=1,
  showstringspaces=false,
  keywordstyle=\color{blue}\bfseries,
  commentstyle=\color{DarkGreen}\itshape,
  breaklines=true,        % <-- wrap long lines
  breakatwhitespace=true, % <-- prefer breaking at whitespace
  postbreak=\mbox{\textcolor{red}{$\hookrightarrow$}\space}, % optional marker
  columns=fullflexible,
  keepspaces=true,
  frame=single,
  caption={Example for existing flaky test in MySQL.}, 
  language=C++,
  captionpos=b, 
  label={lst:nio-example},
  float
]
  TEST(
    ut0new_new_delete_arr,
    destructors_of_successfully_instantiated_
    trivially_constructible_elements_are_invoked
    _when_one_of_the_element_constructors_throws) {
  static int n_constructors = 0;
  static int n_destructors = 0;

  bool exception_thrown_and_caught = false;
  try {
    auto ptr = ut::new_arr_withkey<Type_that_may_throw>(
        ut::make_psi_memory_key(pfs_key), ut::Count(7));
    ASSERT_FALSE(ptr);
  } catch (std::runtime_error &) {
    exception_thrown_and_caught = true;
  }
  EXPECT_TRUE(exception_thrown_and_caught);
  EXPECT_EQ(n_constructors, 4);
  EXPECT_EQ(n_destructors, 3);
}
\end{lstlisting}
\begin{lstlisting}[
  basicstyle=\ttfamily\scriptsize,
  numbers=left,
  numbersep=6pt,
  stepnumber=1,
  showstringspaces=false,
  breaklines=true,        
  breakatwhitespace=true, 
  postbreak=\mbox{\textcolor{red}{$\hookrightarrow$}\space}, 
  keywordstyle=\color{blue}\bfseries,
  commentstyle=\color{DarkGreen}\itshape,
  columns=fullflexible,
  keepspaces=true,
  frame=single,
  language=SQL,
  caption={Example for a flaky test of the unordered collection category generated by GPT-4o for DuckDB.}, 
  captionpos=b, 
  label={lst:unordered-example},
  float
]
-- Test with UNION BY NAME and complex data types
query I
SELECT {'key': 'val'} AS json_col UNION BY NAME SELECT {'key': 123} AS json_col;
----
{'key': val}
{'key': 123}
\end{lstlisting}
\begin{lstlisting}[
  basicstyle=\ttfamily\scriptsize,
  numbers=left,
  numbersep=6pt,
  stepnumber=1,
  showstringspaces=false,
  breaklines=true,        
  breakatwhitespace=true, 
  postbreak=\mbox{\textcolor{red}{$\hookrightarrow$}\space}, 
  keywordstyle=\color{blue}\bfseries,
  commentstyle=\color{DarkGreen}\itshape,
  columns=fullflexible,
  keepspaces=true,
  frame=single,
  language=SQL,
  caption={Example for a flaky test of the random category generated by GPT-4o for DuckDB.}, 
  captionpos=b, 
  label={lst:random-example},
  float
]
-- Test with a large number of random values
statement ok
CREATE TABLE random_values AS
SELECT random() * 1000 - 500 AS x FROM range(10000);
query I
SELECT mad(x) FROM random_values
----
250
\end{lstlisting}

\subsection{RQ2: Root Causes}
We manually inspected the resulting flaky tests to gain insights into the root causes of their flakiness.
We report the results of our manual categorization in \Cref{tab:flaky-issues-counts}. 

Overall, the most prevalent category for flaky LLM-generated tests was \enquote{unordered collection}, appearing in 33 of 52 flaky tests for GPT-4o (65\%) and 38 of 63 tests for Mistral (60\%). All flaky tests in the \enquote{unordered collection} category were generated for DuckDB. In these cases, the LLMs generated SQL statements that assumed an order for the result of a query, without proper use of \texttt{ORDER BY}. \Cref{lst:unordered-example} illustrates an example of such a test, which was generated for a file containing tests for the \texttt{UNION BY NAME} operator. Generally, DuckDB offers two base operators for performing unions, \texttt{UNION} and \texttt{UNION ALL}~\cite{duckdb-blog}. The difference between \texttt{UNION} and \texttt{UNION ALL} is that \texttt{UNION} performs deduplication when stacking, thereby neglecting the original row order. However, in the case of the example in \Cref{lst:unordered-example}, the LLM utilized the \texttt{UNION BY NAME} operator instead of \texttt{UNION ALL BY NAME}, without an explicit \texttt{ORDER BY}. Thus, the test suffered from flaky failures as the values might be returned in the wrong order.

\begin{figure}
    \centering
    \includegraphics[width=0.35\textwidth]{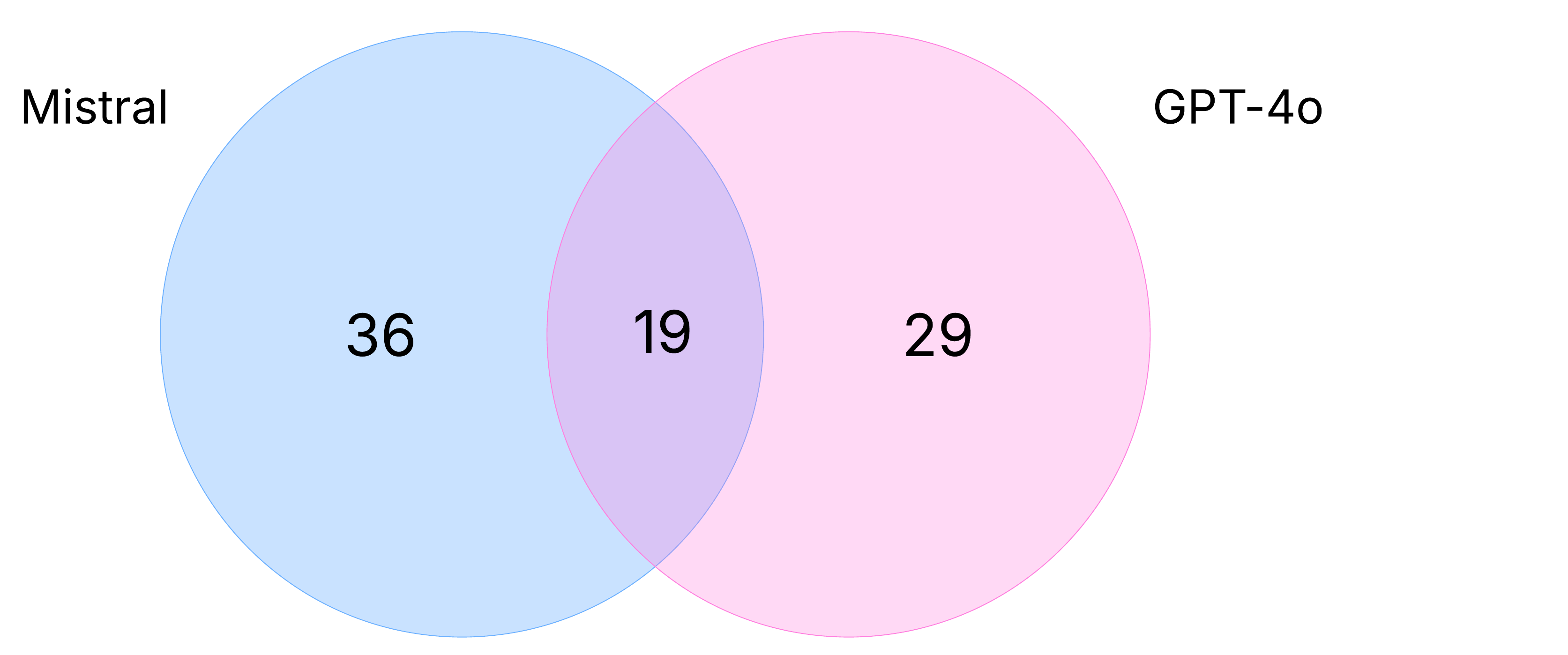}
    \caption{Venn diagram showing the number of files containing generated flaky tests for GPT-4o and Mistral, respectively.}
    \label{fig:venn}
    \Description{A Venn diagram showing there is a notable overlap in files containing flaky tests for both GPT-4o and Mistral.}
\end{figure}

Generally, the distribution of flakiness categories appeared similar between the two models. As shown in \Cref{fig:venn}, there was also a notable overlap in the files that include the generated flaky tests, suggesting that some files are particularly prone to flakiness. \Cref{lst:random-example} shows an example of a test that suffered from randomness, which was generated by both the GPT-4o and the Mistral model for the file \texttt{test\_mad.test}. In addition to \enquote{unordered collection}, we found \enquote{NIO} flakiness to be a problem for both LLMs, appearing in 12 of 115 flaky tests overall (10\%). Nine of these 12 flaky tests were generated for one file in SAP HANA's repository, five by Mistral and four by GPT-4o. In this case, the LLMs employed an existing test function that increases a counter variable without properly cleaning the state at the end of the test. Since this function also causes flakiness in existing tests, the models merely followed the flakiness that was already present in the context.

For tests of the \enquote{concurrency} category, there is a notable difference between the models. Eight out of nine tests that suffered from concurrency were generated by Mistral. These tests were generated for SQLite, MySQL, and DuckDB. For SQLite and DuckDB, they tested the respective thread handling implemented in the project. In DuckDB, the issue was related to the locking functionality. In these cases, similar to the tests in \enquote{unordered collection}, Mistral lacked knowledge of parallelism handling in the respective projects.



\begin{table}
  \caption{Root cause category frequency of flaky tests.}
  \label{tab:flaky-issues-counts}
  \centering
  \small
  \begin{tabular}{p{0.2\textwidth} r r | r}
    \toprule  
    Type & GPT-4o & Mistral & Total \\
    \midrule
    Unordered Collection    & 34 & 38 & 72 \\
    NIO                    & 5  & 7 & 12 \\
    Concurrency            & 1  & 8 & 9 \\
    Unknown                & 4  & 2 & 6 \\
    I/O                    & 2  & 3 & 5 \\
    Randomness             & 1  & 3 & 4 \\
    Other                  & 3  & 0  & 3\\
    Uninitialized Variable & 1  & 1 & 2\\
    Time                   & 1  & 1 & 2\\
    Floating Point         & 0  & 0 & 0\\
    Too Restrictive Range  & 0  & 0 & 0\\
    Test Case Timeout      & 0  & 0 & 0\\
    \midrule
    Total & 52 & 63 & 115\\
    \bottomrule
  \end{tabular}
\end{table}

\begin{table*}
  \caption{The results from our test amplification approach using the test files with injected flaky assertions.}
  \label{tab:injected}
\small
  \centering
    \begin{tabular}{l l r r r r r r}
      \toprule
      & Project & \#Diffs & \#Compilation Successes & \#Added Tests & \#Tests w/ Flaky Assertion & \#Passing Tests & \#Flaky Tests \\
      \midrule
      \multirow{2}{*}{GPT-4o} 
      & SAP HANA & \num{1505} & 638 (42\%) & \num{2794} & 1939 (69\%) & \num{1510} & \num{1052} (70\%) \\
      & MySQL & 259 & 176 (68\%) & 792 & 331 (42\%) & 526 & 217 (41\%) \\
      \midrule
      \multirow{2}{*}{Mistral} 
      & SAP HANA & 1505 & 652 (43\%) & \num{2591} & \num{2012} (78\%) & \num{1467} & \num{1051} (72\%) \\
      & MySQL & 252 & 142 (56\%) & 867 & 507 (58\%) & 715 & 319 (45\%) \\
      \bottomrule
    \end{tabular}
\end{table*}

\textbf{Repeated Executions.} To gain insights into whether the flaky tests persist in repeated executions of our experiment, we repeated our generation for the files containing flaky tests in the context of SAP HANA and MySQL two additional times. We found that every flaky test in the initial iteration was reproduced in at least one of the repetitions for the \cpp~tests. For MySQL, the single flaky test observed in our initial experiment was reproduced in both repetitions. For SAP HANA, we found that one flaky test was only reproduced in a single repetition, whereas the semantic copies of the other four flaky tests were contained in both iterations. 

For the SQL tests, we observed that three of four flaky tests for SQLite were reproduced in two additional repetitions. For DuckDB, of the 42 flaky tests in the initial experiment, we were able to reproduce 37 in the first and 40 in the second repetition.
Overall, we conclude that there was a sufficiently large overlap of the root causes between repetitions.

\begin{framed}
\noindent\textbf{Result RQ2 (Root causes):} Unordered collection was the most prevalent root cause of flaky tests in our study. All instances of tests suffering from this cause were SQL tests, where the LLMs expected an ordered result set without proper use of order by. We also found NIO to be a common root cause of flakiness in LLM-generated tests, mostly caused by LLM following the flakiness of existing tests. 
\end{framed}

\subsection{RQ3: Flakiness Transfer}
We report the results of our flakiness transfer experiment in \Cref{tab:injected}.
As shown in \Cref{tab:injected}, both models were susceptible to transferring the injected assertion to the newly generated tests. In addition, GPT-4o and Mistral showed an increased susceptibility to transferring the flaky assertion in the context of SAP HANA compared to MySQL. The proportion of tests containing the flaky assertion was approximately 20\% higher between projects for both models. We conjecture that this is due to the background knowledge the models have gained on MySQL during pre-training. 

Comparing between models, GPT-4o yielded a lower proportion of tests containing the flaky assertion by a factor of 0.88 for SAP HANA and 0.72 for MySQL. 
Looking at the difference between passing and non-passing tests, there was only a negligible difference between the proportion of tests containing the flaky assertion for GPT-4o. For tests generated by Mistral, the proportion was lower by a factor of 0.92 for SAP HANA and 0.78 for MySQL. On a file-level, we observed that the LLM most often either transfers the flaky assertion to all newly generated tests in a file, or to none of them. For example, the \num{1052} resulting flaky tests for SAP HANA generated by GPT-4o arose from 344 generated diffs, i.e., 23\% of all generated diffs. However, only one of these diffs contained both tests with and without the flaky assertion. Similarly, the flaky tests generated for SAP HANA by Mistral are contained in two files.

\textbf{Repeated Executions.} Based on two additional repetitions of our experiments, we observed only marginal changes compared to the results in \Cref{tab:injected}, with the largest deviation being 2\%.
For MySQL, the proportion of tests that contained the flaky assertion in our repetitions was 40\% and 42\% compared to 42\% in our initial experiment.
For SAP HANA, we observed 68\% and 71\% compared to 69\% in our initial experiment.

\begin{framed}
\noindent\textbf{Result RQ3 (Following):} Both LLMs were susceptible to transferring the flakiness present in the provided context. Compared to GPT-4o, a higher proportion of tests generated by Mistral contained the flaky assertion. For both models, the prevalence was higher in the context of SAP HANA compared to MySQL. 
\end{framed}

\section{Discussion}
\label{sec:discussion}
In this section, we discuss the implications of our empirical results.

\textbf{On LLM-based Test Generation.}
Overall, we observed that the LLM-based test generation approach of \citet{meta-generation} resulted in a reasonable number of generated tests.
Although we chose the simplest possible setup to generate tests using an LLM, we were able to generate compilable diffs, even for complex projects such as SAP HANA.
However, compared to the original study of \citet{meta-generation}, we observed a notable decrease in compilation success.
This decrease is expected due to the inherent complexity of database systems and the complexity of \cpp~compared to Kotlin.
The drop in compilation success for SAP HANA confirms our assumption that LLMs may yield worse results in a large and complex closed-source project.
Based on the results of our study, future work at SAP is currently exploring approaches to provide LLMs with the required context to handle the large number of custom libraries used within SAP HANA's source code, which are not known to public LLMs. We view the simple setup of this study as a starting point for future work on this topic, which may employ more sophisticated approaches to test generation. For example, future work could experiment with other setups than test amplification or investigate the impact of different prompting techniques.

\textbf{On the Resulting Proportion of Flaky Tests.}
We found that both LLMs generate flaky tests across all projects. 
The resulting proportion of flaky tests for LLM-generated \cpp~tests was slightly higher than the proportion of flaky tests in the existing tests.
Consequently, we expect that the use of LLMs for the test generation could increase the overall proportion of flaky tests in a test suite. 
Furthermore, fixing flaky tests is often neglected due to other priorities~\cite{hoang2024presubmit,berndt2024taming}, even for existing tests. 
For generated tests, the lack of ownership for these tests may exacerbate this effect, leaving LLM-generated flaky tests as a long-term liability in the test suite.

\textbf{On the Results of our Manual Inspection.} Based on our manual inspection of flaky tests, we conjecture that a lack of context on the system under test may be the major cause of flakiness in LLM-generated tests. This seems particularly pronounced for SQL tests, where the LLM had no information on when the system under test guarantees the order of query results.
While adding related context may be a valid solution in the context of unit tests, where the test's scope is by definition limited, such approaches are challenging to apply to SQL tests, as they may require a large amount of context.
For example, with millions of SLOC, SAP HANA's source code does not fit entirely into the context windows of contemporary LLMs.
Therefore, we encourage future work on approaches that automatically provide LLMs with tailored context for generating system tests, which requires information about the entire software. 

\textbf{On the Impact of Flakiness Injection.}
Overall, we observe that the LLM transfers the injected flaky assertion to a notable number of newly generated tests for both study subjects.
Since the flaky assertion in the provided context is clearly not causally related to the task at hand, we attribute this finding to the LLM's susceptibility to spurious correlations in in-context learning. We conjecture that the LLM overvalues the provided context compared to the knowledge gained from pre-training.
While this may lead to desirable effects such as the \enquote{Fashion Following} of existing coding conventions~\cite{harman2025harden}, the results of our simple setup suggest that this happens without any semantic understanding of the given code, which may also lead to undesired properties in code synthesized by LLM in practice.
Therefore, we encourage future work to investigate this effect. For example, future studies could further validate the degree to which LLMs propagate non-functional code properties, such as code smells, from the code shown for in-context learning.

\section{Conclusions}
\label{ch:conclusion}
To gain insights into the flakiness of LLM-generated tests for database systems, we used two LLMs to generate two types of tests commonly encountered in database systems: native \cpp~unit tests and SQL tests that run SQL statements.
We applied an approach for LLM-based test amplification to three open-source databases and SAP HANA, a large industrial database, and examined the flakiness of the resulting tests. 

Our study revealed that LLMs struggle to generate compilable \cpp~source code, especially in complex closed-source projects such as SAP HANA. 
Our analysis showed that the most common root cause of flakiness generated by LLMs in the context of our study was SQL statements in which the LLM expects a certain order without proper use of ORDER BY.
We found that the proportion of flaky tests as well as root causes varied between LLMs.
However, both LLMs transferred existing flakiness from existing tests in the given context of a prompt to newly generated tests.
We attribute this problem to the shortcut learning of LLMs, where they follow spurious correlations in the provided context during in-context learning. To further understand this phenomenon, we injected a flaky assertion into all existing tests before repeating our generation experiment with the injected files.
Our results suggest that LLMs are susceptible to transferring the injected assertion.
We found that both LLMs were more susceptible to our injection in the context of SAP HANA compared to MySQL.
We attribute this finding to the fact that public LLMs lack prior knowledge of SAP HANA, as its source code was not part of their training data, leading to greater reliance on the given context in the prompt.

Our findings substantiate the importance of providing LLMs with appropriate context when employing them for test generation. We motivate practitioners to pay attention not only to engineering goals such as coverage or mutation scores when evaluating LLMs for test generation, but also to consider potential side effects of the resulting tests, such as flakiness.
We encourage future studies to evaluate and extend our test generation setup by including additional test generation approaches and models.
We also encourage researchers to investigate the impact of shortcut learning on the application of LLMs for automating software engineering tasks.

\balance
\bibliographystyle{ACM-Reference-Format}
\bibliography{bibliography}

\end{document}